\documentclass[aps,pra,superscriptaddress,showpacs,twocolumn]{revtex4-1}
\usepackage{graphicx,bm,amsmath,amsfonts,bbold, amssymb,color,microtype}

\newcommand{\be}{\begin{equation}}
\newcommand{\ee}{\end{equation}}
\newcommand{\bea}{\begin{eqnarray}}
\newcommand{\eea}{\end{eqnarray}}
\newcommand{\im}[1]{\mbox{Im}[#1]}
\newcommand{\re}[1]{\mbox{Re}[#1]}

\newcommand{\tauEP}{\tau_{\!\scriptscriptstyle \text{TH}}}

\newcommand{\dz}[1]{\frac{d #1}{dz}}
\begin{document}

\title{Optical fluxes in coupled $\cal PT$-symmetric photonic structures}

\author{Li Ge}
\email{li.ge@csi.cuny.edu}
\affiliation{\textls[-18]{Department of Engineering Science and Physics, College of Staten Island, CUNY, Staten Island, NY 10314, USA}}
\affiliation{The Graduate Center, CUNY, New York, NY 10016, USA}

\author{Konstantinos G. Makris}
\affiliation{Department of Physics, University of Crete, P.O. Box 2208, 71003, Heraklion, Greece}

\author{Lingxuan Zhang}
\affiliation{\textls[-18]{Department of Engineering Science and Physics, College of Staten Island, CUNY, Staten Island, NY 10314, USA}}
\affiliation{\textls[-40]{State Key Laboratory of Transient Optics and Photonics, Xi'an Institute of Optics and Precision Mechanics, Chinese Academy of Sciences, Xi'an 710119, China}}

\begin{abstract}
In this work we first examine transverse and longitudinal fluxes in a $\cal PT$-symmetric photonic dimer using a coupled-mode theory.
Several surprising understandings are obtained from this perspective: The longitudinal flux shows that the $\cal PT$ transition in a dimer can be regarded as a classical effect, despite its analogy to $\cal PT$-symmetric quantum mechanics. The longitudinal flux also indicates that the so-called giant amplification in the $\cal PT$-symmetric phase is a sub-exponential behavior and does not outperform a single gain waveguide. The transverse flux, on the other hand, reveals that the apparent power oscillations between the gain and loss waveguides in the $\cal PT$-symmetric phase can be deceiving in certain cases, where the transverse power transfer is in fact unidirectional. We also show that this power transfer cannot be arbitrarily fast even when the exceptional point is approached. Finally, we go beyond the coupled-mode theory by using the paraxial wave equation and also extend our discussions to a $\cal PT$ diamond and a one-dimensional periodic lattice.
\end{abstract}


\maketitle

\section{introduction}

Parity-Time ($\cal PT$) symmetric optical systems have attracted growing interest in the past several years \cite{El-Ganainy_OL07,Moiseyev,Musslimani_prl08,Makris_prl08,Kottos,Guo,Makris_PRA10,mostafazadeh,Longhi,CPALaser,conservation,conser2D,Robin,EP9,nonlinearPT,Ruter,Lin,Feng_NM,Feng2,Walk,Hodaei,Yang,Chang,Ramezani}. These systems are non-Hermitian due to the presence of gain and loss, which are delicately balanced such that the refractive index satisfies $n(x)=n^*(-x)$ with respect to a chosen symmetry plane at $x=0$. A plethora of findings in such systems are tied to the spontaneous $\cal PT$ symmetry breaking at an exceptional point (EP) \cite{EP1,EP2,EPMVB,EP3,EP4,EP5,EP6,EP8}, which was first investigated theoretically in non-Hermitian quantum mechanics \cite{Bender1,Bender2,Bender3} and later realized in photonics in the paraxial regime \cite{El-Ganainy_OL07,Musslimani_prl08,Makris_prl08,Guo,Ruter,Makris_PRA10}. The energy levels of the system (or propagation constants for waveguides) undergo a transition from real to complex conjugate pairs as a result of this spontaneous symmetry breaking.

The study of paraxial wave propagation has focused on the eigenvalue spectra and wave dynamics. In contrast, optical fluxes in both the longitudinal and transverse directions have not received much attention \cite{Konotop_flux} especially when the system is not in a propagation eigenstate.
They are important quantities as shown in a different $\cal PT$-symmetric setup, i.e., light scattering from a $\cal PT$-symmetric structure, where a conserved flux between the incident and scattered waves marks the $\cal PT$-symmetric phase of the scattering matrix \cite{CPALaser,Robin}. A generalized conservation law in terms of transmittance and reflectance was also found for both one-dimensional (1D) and quasi-1D $\cal PT$-symmetric structures \cite{conservation,conser2D}, which holds independent of the $\cal PT$ transition.

In this work we examine longitudinal and transverse optical fluxes in $\cal PT$-symmetric photonic systems, starting with a $\cal PT$ dimer consisting of two parallel waveguides in the coupled-mode regime. The longitudinal flux reflects the change of the total intensity as a function of the propagation distance (time) and is hence a measure of local gain and loss. The transverse flux, on the other hand, is the result of frustrated total internal reflection through waveguide sidewalls and analogous to quantum tunneling. We found that these two quantities in fact reveal several important properties of previous studied phenomena.

More specifically, by considering the longitudinal flux we show that the standard $\cal PT$ transition in a photonic dimer, where the propagation constants change from real to a complex conjugate pair, can be regarded as a classical effect: it depends only on the intensity pattern of a propagation eigenmode but not on the relative phase of its amplitudes in the two waveguides. The longitudinal flux also indicates that the so-called giant amplification in the $\cal PT$-symmetric phase \cite{Konotop} is a sub-exponential behavior and does not outperform a single gain waveguide.

Using the transverse flux, we analyze transverse power oscillations in the $\cal PT$-symmetric phase \cite{Makris_prl08,El-Ganainy_OL07,Ruter}, which was among the first theoretical predications that have promoted the fast growing interests in $\cal PT$-symmetric optics. A similar effect was also found in a complex Su-Schrieffer-Heeger (SSH) lattice \cite{Ramezani}.
One often overlooked property of this behavior is that the distance (time) of the oscillation from the gain waveguide to the loss waveguide is shorter than the reversed process. We reveal that the apparent power transfer from the gain waveguide to the loss waveguide cannot be arbitrarily fast as the system approaches the EP; it is bounded by the inverse of the coupling constant. Interestingly, although the transverse flux is bidirectional in general and hence consistent with the apparent shifting of the intensity peak from one waveguide to the other, we show that these transverse oscillations can be deceiving in certain cases, where the transverse flux is in fact \textit{unidirectional}.

Finally, we go beyond the coupled-mode theory and confirm the aforementioned findings using the paraxial wave equation. We also extend our discussion to a $\cal PT$-symmetric ``diamond", where the gain and loss waveguides are coupled indirectly by two separate bridging waveguides without gain or loss. Despite the coincidence of a genuine degeneracy and an EP of order 3 \cite{Graefe}, the power oscillations persist and continue to show unidirectional character in some cases. These behaviors are also expected in certain more complicated $\cal PT$-symmetric photonic structures, such as a one-dimensional lattice formed by coupling many identical $\cal PT$ dimers.

\section{Transverse and longitudinal fluxes in a Hermitian dimer}

We start our discussion by briefly reviewing the transverse and longitudinal fluxes in a Hermitian dimer. We use the following coupled-mode theory in the paraxial regime
\begin{align}
i\frac{\partial a}{\partial z} &= \beta a + J b, \label{eq:a}\\
i\frac{\partial b}{\partial z} &= \beta b + J a, \label{eq:b}
\end{align}
where $z$ is the propagation distance, $\beta$ is the propagation constant, and $J$ is the coupling coefficient. Let us rewrite the two equations above in terms of intensities $I_a\equiv|a|^2$ and $I_b\equiv|b|^2$:
\begin{align}
\frac{\partial I_a}{\partial z} &= - iJ(a^*b-ab^*), \label{eq:I_a}\\
\frac{\partial I_b}{\partial z} &= iJ(a^*b-ab^*). \label{eq:I_b}
\end{align}
The Hermitian nature of the system is reflect by the fact that the longitudinal flux, defined by ${\cal J}_z \equiv \partial_z (I_a+I_b)$, vanishes. Consequently, the transverse flux ${\cal J}_t\equiv  iJ(a^*b-ab^*)$ leaving waveguide $a$ all enters waveguide $b$. This example indicates that the magnitudes of the fluxes in the longitudinal and transverse directions are not necessarily correlated. We also note that ${\cal J}_t$ does not depend on the propagation constant $\beta$, and it is equivalent to $L_y$ in the angular momentum representation of the effective Hamiltonian \cite{Graefe2,Ramezani2}.

To study the dynamics of ${\cal J}_t$, one can derive the $z$-derivative of $a^*b$ directly using Eqs.~(\ref{eq:a}) and (\ref{eq:b}). Here we take a different approach and utilize Eqs.~(\ref{eq:I_a}) and (\ref{eq:I_b}) instead, with which we rewrite ${\cal J}_t$ as
\be
{\cal J}_t = \frac{1}{2}\frac{\partial \Delta }{\partial z},\quad\Delta\equiv I_a-I_b.
\ee
By differentiating Eqs.~(\ref{eq:a}) and (\ref{eq:b}) in the rotating frame ($a,b\rightarrow ae^{-i\beta z},be^{-i\beta z}$), it is straightforward to find
\be
\frac{\partial I_{a,b}}{\partial z} + 4J^2 I_{a,b}=0,
\ee
which gives us
\be
\Delta(z) = \Delta_s \sin(2Jz) + \Delta_c \cos(2Jz).
\ee
$\Delta_{s,c}$ are two real constants determined by the initial conditions $\Delta(0)$ and $\partial_z\Delta(0)$. As a result, we find
\be
{\cal J}_t(z) = J [\Delta_s \cos(2Jz) - \Delta_c \sin(2Jz)].
\ee
Clearly ${\cal J}_t(z)$ is a sinusoidal function of spatial period $\pi/J$, and it resembles Rabi oscillations of a two-level system in quantum mechanics. Here the (transverse) power oscillations are symmetric, i.e., it takes the same propagation distance for power to oscillate back and forth between the two waveguides.

Alternatively, one can analyze the wave dynamics by decomposing the wave function in the eigenstate basis, i.e.,
\be
\Psi(z) \equiv \begin{pmatrix} a \\ b \end{pmatrix} = c_+\Psi_+e^{-i\lambda_+z} + c_-\Psi_-e^{-i\lambda_-z},\label{eq:decomp}
\ee
where $\Psi_\pm = [1,\, \pm1]^T$ and the corresponding eigenvalues are given by $\lambda_\pm = \beta\pm J$. The superscript $T$ denotes the matrix transpose, and $c_\pm$ are the expansion coefficients at $z=0$, which are complex in general. Denoting $\theta = \text{Arg}[c_+]-\text{Arg}[c_-]$, $I_a$ then reaches its maximum at $z=(\theta + 2n\pi)/2J\,(n=0,1,2,\ldots)$. Similarly, $I_b$ is maximized at $z=[\theta + (2n+1)\pi]/2J$, and it takes the same distance $z=\pi/2J$ for power to oscillate completely from one waveguide to the other, which agrees with the result derived above.

\section{Fluxes in a $\cal PT$ dimer}
\label{sec:PTdimer_CMT}

Next we turn to a $\cal PT$ dimer with equal gain and loss:
\begin{align}
i\frac{\partial a}{\partial z} &= (\beta + i\tau ) a + J b, \label{eq:a_pt}\\
i\frac{\partial b}{\partial z} &= (\beta - i\tau ) b + J a. \label{eq:b_pt}
\end{align}
The equations for the intensities in the two waveguides are now given by
\begin{align}
\frac{\partial I_a}{\partial z} &= 2\tau I_a - {\cal J}_t, \label{eq:I_a_pt}\\
\frac{\partial I_b}{\partial z} &= -2\tau I_b + {\cal J}_t. \label{eq:I_b_pt}
\end{align}
Clearly the terms proportional to $\tau$ are due to intra-waveguide gain and loss, and the transverse flux is still given by ${\cal J}_t=iJ(a^*b-ab^*)$ while the longitudinal flux is now given by ${\cal J}_z = 2\tau(I_a-I_b)$.

\subsection{``Classical" $\cal PT$ transition}
We note that in an eigenstate $\Psi(z)=\Psi(0)e^{-i\lambda z}$, which indicates $\im{\lambda} = {\cal J}_z/2(I_a+I_b)$. The two phases of $\cal PT$ symmetry can then be judged by combing this expression with ${\cal J}_z = 2\tau(I_a-I_b)$ given above: in one phase $I_a=I_b$ and $\lambda$ is real, resulting in a conserved longitudinal flux; in the other $I_a\neq I_b$ and $\lambda$ is complex, and the resulting positive (negative) ${\cal J}_z$ indicates wave amplification (attenuation). Since the relative phase between the complex amplitudes $a,b$ does not appear in this criterion, we can classify this standard $\cal PT$ transition as a ``classical" one. In contrast, this relative phase drives the anomalous nonlinear $\cal PT$ transition \cite{Anomalous} and pseudo-Hermitian transition \cite{FWM}, which can be retreated as ``quantum" transitions in this regard. This distinction based on whether the relative phase of two complex amplitudes appears is familiar in the double-slit experiment, the outcome of which differs for classical particles and quantum particles.

\subsection{Asymmetric power oscillations and unidirectional evanescent flux}

\begin{figure}[t]
\centering
\includegraphics[clip,width=\linewidth]{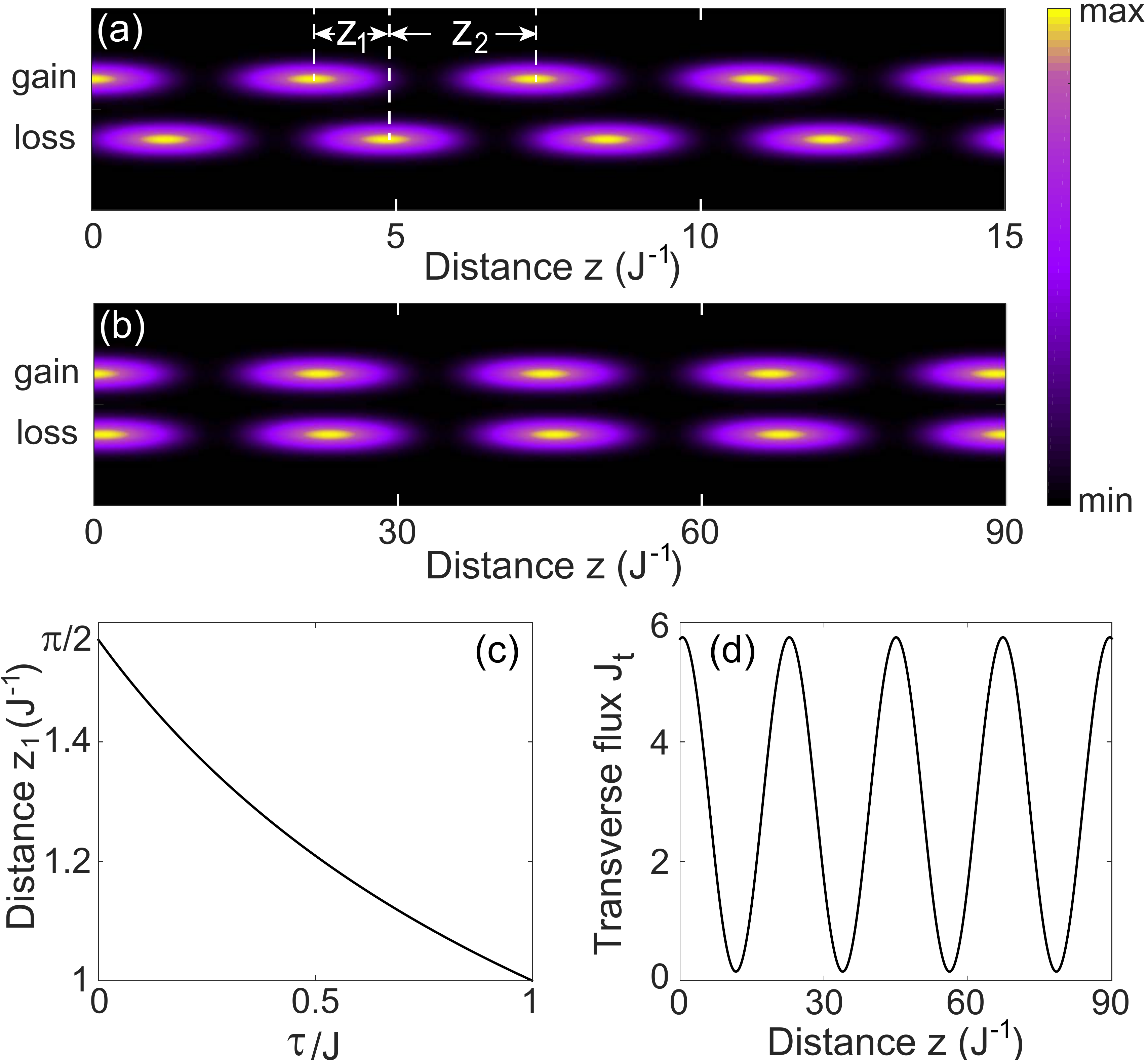}
\caption{(Color online) Asymmetric power oscillations in a $\cal PT$ dimer. (a) and (b) Intensities $I_{a,b}$ in the two waveguides, where $c_+=1,\,c_-=0.7$ and $\tau/J=0.5$ and 0.99, respectively. The transverse flux in (b) is shown in (d).
(c) $z_1$ as a function of $\tau/J$.} \label{fig:osc}
\end{figure}

Equations~(\ref{eq:I_a_pt}) and (\ref{eq:I_b_pt}) are different from their Hermitian counterparts [Eqs.~(\ref{eq:I_a}) and (\ref{eq:I_b})] even in the $\cal PT$-symmetric phase, which indicates that the expectation of Rabi-like oscillations in a $\cal PT$ dimer based on its real energy eigenvalues  is problematic. In fact, these equations already show clearly that the transverse power oscillations are asymmetric: when the intensity $I_a$ in the gain waveguide is at its maximum (and hence larger than the intensity $I_b$ in the loss waveguide), Eq.~(\ref{eq:I_a_pt}) tells us that the transverse flux ${\cal J}_t$ is positive and equal to $2\tau I_a$; hence it is larger than $2\tau I_b$ and leads to a positive $\partial_z I_b$ using Eq.~(\ref{eq:I_b_pt}), meaning that $I_b$ is in the half period to reach its maximum. Similarly, when $I_b$ is at its maximum, we know that $I_a$ is in the half period to reach its minimum [see Figs.~\ref{fig:osc}(a) and \ref{fig:osc}(b)].

These asymmetric behaviors are often overlooked in the discussion of the power oscillations. To understand them more quantitatively, we again decompose the wave function in the eigenstate basis using Eq.~(\ref{eq:decomp}). The eigenstates are now given by $\Psi_\pm = [1, \pm e^{\mp i\phi}]$ and the corresponding eigenvalues are $\lambda_\pm=\beta\pm J\cos\phi$, where $\phi\equiv\arcsin(\tau/J)\in(0,\,\pi/2)$ in the $\cal PT$-symmetric phase. To simplify the notation, we choose $z=0$ to be the point where the expansion coefficients $c_\pm$ are in phase, and we denote $c_+=rc_- (r>0)$ for a non-eigenstate in the $\cal PT$-symmetric phase. The intensities in the gain and loss waveguides are then given by
\begin{align}
I_a(z) &= |c_-|^2[1+r^2+2r\cos2\phi_1(z)],\label{eq:I_a_pt2}\\
I_b(z) &= |c_-|^2\{1+r^2+2r\cos[2\phi_1(z)-(\pi-2\phi)]\},\label{eq:I_b_pt2}
\end{align}
where $\phi_1(z) \equiv Jz\cos\phi$. $I_{a,b}(z)$ undergo sinusoidal oscillations, but they are $\pi-2\phi$ out of phase. Therefore, the distance $z_{1} = (\pi-2\phi)/(2J\cos\phi)$ between one peak of $I_a$ and the subsequent peak of $I_b$ is shorter than the distance $z_{2} = (\pi+2\phi)/(2J\cos\phi)$ between this peak of $I_b$ and the next peak of $I_a$, leading to an asymmetric power oscillation [see Fig.~\ref{fig:osc}(a)]. 
This behavior is very different from Rabi oscillations in a Hermitian dimer (i.e., when $\phi=0$), and it highlights an important impact of non-Hermiticity, i.e., the eigenmodes of the system are no longer orthogonal, even when the eigenvalues are still real.

As we approach the EP where $\tau=J$ and $\phi=\pi/2$, $I_{a,b}$ oscillate more and more in phase [see Fig.~\ref{fig:osc}(b)].
Despite this trend, it is important to note that the apparent power transfer from the gain waveguide to the loss waveguide does not become arbitrarily fast as the system approaches the EP. To be more precise, $z_1$ reduces monotonically with $\tau/J$ but is bounded by $J^{-1}$ from below [see Fig.~\ref{fig:osc}(c)]. This behavior can be understood in the following way: although the ratio $z_1/z_2$ approaches zero as the system approaches the EP, the spatial period of $I_{a,b}(z)$ [i.e., $z_0=z_1+z_2\equiv\pi/(J\cos\phi)$] approaches infinity as well, which leads to a finite value of $z_1$ but leaves $z_2$ diverging. To show these observations quantitatively, we define $\delta=J-\tau$ and $\theta=\pi/2-\phi$. It then follows that $\cos\theta=1-\delta/J\approx1-\theta^2/2$ as we approach the EP, or equivalently, $\theta\approx\sqrt{2\delta/J}$. Therefore, we find
\begin{align}
z_1 &= \frac{\pi-2\phi}{2J\cos\phi} \approx \frac{2\theta}{2\sqrt{2J\delta}} = J^{-1}, \label{eq:z1}\\
z_2 &= \frac{\pi+2\phi}{2J\cos\phi} \approx \frac{2\pi}{2\sqrt{2J\delta}} \rightarrow \infty. \label{eq:z2}
\end{align}

So far we have discussed asymmetric transverse power oscillations without actually calculating how much power is directly transferred between the waveguides. We find the transverse flux ${\cal J}_t$ in the $\cal PT$-symmetric phase to be
\be
{\cal J}_t(z) = 2|c_-|^2\{(1+r^2)\tau + 2rJ\sin[2\phi_1(z)+\phi]\}
\ee
using Eqs. (\ref{eq:I_a_pt}) and (\ref{eq:I_a_pt2}) [or Eqs. (\ref{eq:I_b_pt}) and (\ref{eq:I_b_pt2})]. In order for the power oscillations between the two waveguides to be symmetric, i.e., a sinusoidal function with a zero mean, the offset $2|c_-|^2(1+r^2)\tau=2(|c_+|^2+|c_-|^2)\tau$ from zero in the expression above needs to vanish, which holds only in the Hermitian case (i.e., $\tau=0$). Interestingly, ${\cal J}_t(z)$ can flow \textit{unidirectionally} from the gain waveguide to the loss waveguide, and the apparent transverse power oscillation is merely a disguise. This situation takes place when
\be
\tau>\frac{2Jr}{1+r^2},\label{eq:uni_criterion}
\ee
which can be satisfied in the $\cal PT$-symmetric phase as long as $r\neq1$. The example shown in Fig.~\ref{fig:osc}(b) is one such case, and its unidirectional ${\cal J}_t$ is shown in Fig.~\ref{fig:osc}(d). Different from previous findings of unidirectional wave propagation in $\cal PT$-symmetric photonic structures \cite{conservation,Lin,Feng_NM}, we note that here the flux is carried by frustrated evanescent waves instead of propagating waves.

\subsection{Giant amplification}

\begin{figure}[b]
\centering
\includegraphics[clip,width=\linewidth]{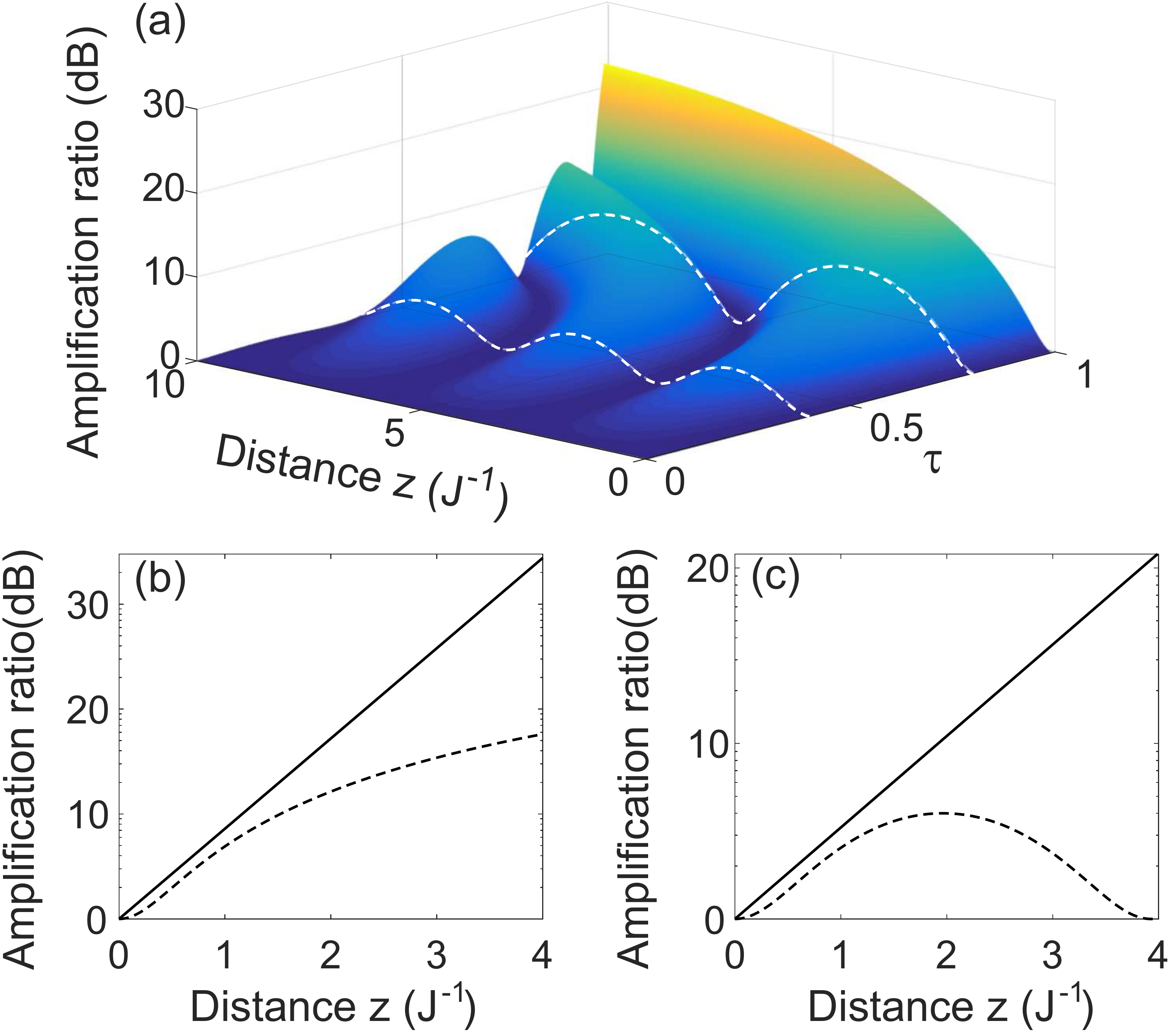}
\caption{(Color online) The values of ``giant amplification" for $\tau/J$ up to 0.99 and $z$ up to $10J^{-1}$. Those at $\tau/J=0.4$ and 0.8 highlighted by white dashed lines. (b)-(c) ``Giant amplification" (dashed lines) compared with the exponential amplification in a single gain waveguide (solid lines). $\tau/J=0.99$ in (b) and 0.5 in (c).} \label{fig:amp}
\end{figure}

The diverging spatial period $z_0$ of power oscillations mentioned in the previous section also needs to be considered when discussing the ``giant amplification" near the EP \cite{Konotop}. To understand this behavior in simple terms, let us consider an equal amplitude ($r=1$) superposition of $\Psi_\pm$. The total intensity can be calculated using the sum of Eqs.~(\ref{eq:I_a_pt2}) and (\ref{eq:I_b_pt2}):
\be
I(z) = 4|c|^2\left\{1+\frac{\tau}{J}\sin[2\phi_1(z)+\phi]\right\}.
\ee
Its minimum $I_\text{min}=4|c|^2(1-\tau/J)$ at $\phi_1(z)=-\phi/2-\pi/4$ and its maximum $I_\text{max}=4|c|^2(1+\tau/J)$ at $\phi_1(z)=-\phi/2+\pi/4$ differ dramatically near the $\cal PT$ transition point, where $I_\text{min}\rightarrow 0$ and $I_\text{max}\rightarrow 8|c|^2$. Now if the input wave is prepared with $I = I_\text{min}$, then one can expect a giant amplification when the intensity reaches $I_\text{max}$. We note that this giant amplification is in fact an interference effect of the two non-orthogonal eigenstates: the minimum intensity is a result of their maximum destructive inference, while the maximum intensity is a result of their maximum constructive interference. The contrast of $I_\text{min}$ and $I_\text{max}$ increases near the EP because the two eigenstates of the system become almost identical. We plot this amplification ratio as a function of propagation distance and $\tau$ in Fig.~\ref{fig:amp}(a).

One concern though is that this giant amplification takes place over the distance $z_0/2$ that also approaches infinity near the EP. Therefore, the important quantity here is the amplification rate \textit{per unit propagation distance}, which is captured by the ratio of the longitudinal flux and the input intensity:
\be
\frac{{\cal J}_z(z)}{I_\text{min}} = 2\tau\sin[2\phi_1(z)]\left(\frac{J+\tau}{J-\tau}\right)^{\frac{1}{2}}.\label{eq:giant_amp}
\ee
Here the point $z=0$ has been shifted from the convention used in the previous section to where the total intensity is at its minimum.
While this expression looks to be divergent near the EP, its value shortly after the launch of the wave, i.e.,
\be
\frac{{\cal J}_z(z)}{I_\text{min}} \approx 4\tau(J+\tau) z\quad(z\ll z_0),
\ee
is actually quite limited in comparison with a single gain waveguide, where $\tilde{I}=I_\text{min}\exp(2\tau z)$ and
\be
\frac{\tilde{\cal J}_z(z)}{I_\text{min}} = 2\tau\exp(2\tau z).
\ee
In fact, one can show that the longitudinal flux in a single gain waveguide is always stronger than that in giant amplification, independent of the value of $z$ after the launch of the wave. To prove this claim, we show that
\be
D(z) = \ln\frac{\tilde{\cal J}_z(z)}{{\cal J}_z(z)}>0.
\ee
Since $J_z(z)$ is periodic while $\tilde{J}_z(z)$ increases exponentially, we only need to prove this inequality in the first period of $J_z(z)$ where $z<z_0$.
We immediately find that $D(z\rightarrow 0)=\ln(2\tau/0^+)$ obviously satisfies this condition, and so does the minimum of $D(z)$, i.e.,
\be
\min[D(z)]
=  \left(\frac{\pi}{2}-\phi\right)\tan\phi - \ln(1+\sin\phi),
\ee
which occurs at $z = (\pi/2-\phi)/(2J\cos\phi)$. Using $x>\sin x$ for $x>0$, or its variant $(\pi/2-\phi)>\cos\phi$ for $\phi\in(0,\pi/2)$ in the $\cal PT$-symmetric phase, we finally derive
\be
\min[D(z)]>\sin\phi-\ln(1+\sin\phi)>0.
\ee
In the last step we have also used $e^x>(1+x)$ for $x>0$.

Therefore, we see that the giant amplification, though interesting as a non-Hermitian effect by itself, does not outperform a single gain waveguide with the same value of $\tau$ (see the examples in Fig.~\ref{fig:amp}). We also note that this giant amplification requires input in both waveguides in general; single-waveguide input implies $\phi_1=\pi/2$ or $\pi-\phi$ from Eqs.~(\ref{eq:I_a_pt2}) and (\ref{eq:I_b_pt2}) when $r=1$, while the input with $I_\text{min}$ requires $\phi_1=-\phi/2-\pi/4$ instead as mentioned previously. Therefore, with single-waveguide input there are alternate regions of amplification and attenuation as the wave propagates, whose period approaches infinity near the EP. In more complicated non-Hermitian structures, the maximum amplification and the corresponding input can be calculated using the singular value decomposition \cite{Makris_PRX14}.

\section{$\cal PT$ dimer in the continuous regime}

Below we extend our discussion of the $\cal PT$ dimer beyond the coupled-mode theory. For simplicity we consider a planar system, where the waveguides are assumed to be invariant along the $z$-axis and characterized by a complex index of refraction that is $\cal PT$-symmetric along the transverse direction, i.e., $n(x)=n^{*}(-x)$. Under the paraxial approximation, the dynamics of the slowing-varying field amplitude $\Phi(x,z)$ is captured by a Schr\"{o}dinger-like equation:
\be
 i\frac{\partial \Phi}{\partial z}=-\frac{\partial^2 \Phi}{\partial^2 x}+V(x)\Phi \equiv H \Phi.
\ee
Note that $x$ and $z$ are in their scaled units and dimensionless. $V(x)$ is the optical potential given by $V(x) \equiv -n_R(x)-i n_I(x)$, where $n_{R,I}(x)$ are the real and imaginary parts of the index modulation on top of a uniform background index. We choose $n_I$ to be positive (negative) for loss (gain). As we mentioned before, the eigenmodes in a non-Hermitian system (here they are functions of the $x$ coordinate only) are non-orthogonal and the eigenvalues $\lambda_n = \beta_n + i\gamma_n$ are complex in general.

\begin{figure}[t]
\centering
\includegraphics[clip,width=\linewidth]{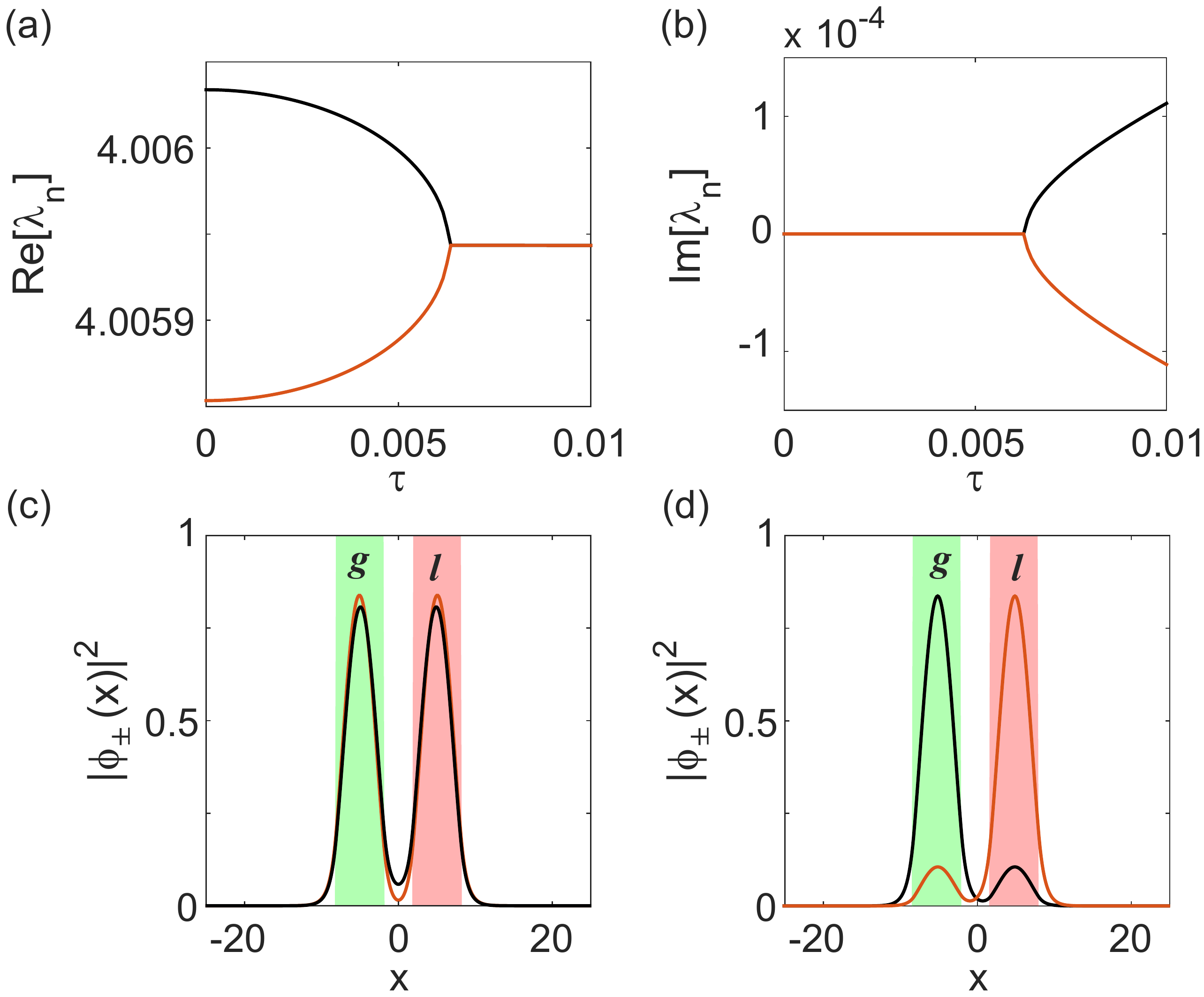}
\caption{(Color online) ``Classical" $\cal PT$ transition in a $\cal PT$ dimer. (a) and (b) Real and imaginary parts of the two complete propagation constant $\lambda_\pm$. (c) and (d) Spatial profiles of these two guided modes at $\tau=0.005$ and $0.01$, respectively. The two waveguides with gain (``$g$") and loss (``$l$") are marked by the shaded regions, with width $w=6$ and center-to-center distance $2d=10$.}\label{fig:transition_cont}
\end{figure}

Let us follow the same order of discussions in Sec.~\ref{sec:PTdimer_CMT} and discuss the ``classical" $\cal PT$ transition first. For the guided modes in the photonic structure, they have a vanishing amplitude as $x\rightarrow\infty$. As a result, it is straightforward to show that the longitudinal flux $J_z\equiv d I(z)/dz$ is given by
\be
J_z = -2\int_{-\infty}^{\infty} dx \, n_I(x) |\Phi(x,z)|^2,\label{eq:P_cont}
\ee
where $I(z)\equiv \int_{-\infty}^{\infty} dx \, |\Phi(x,z)|^2$. Now for an eigenmode $\Phi(x,z)\propto \phi_n(x)e^{-i\lambda_nz}$, $J_z$ becomes $2\gamma_nI$. In the $\cal PT$-symmetric phase we have $|\phi_n(x)|^2=|\phi_n(-x)|^2$ [see Fig.~\ref{fig:transition_cont}(c)], and as a result the right hand side in Eq.~(\ref{eq:P_cont}) vanishes because the integrand is an odd function about $x=0$, leading to a vanished $\gamma_n$ and a real $\lambda_n$ [see Fig.~\ref{fig:transition_cont}(b)]. The situation is different when $|\phi_n(x)|^2\neq|\phi_n(-x)|^2$ in the $\cal PT$ broken phase [see Fig.~\ref{fig:transition_cont}(d)], where $\lambda_n$ becomes complex and Eq.~(\ref{eq:P_cont}) indicates $\gamma_n=-\gamma_{n'}\neq0$ [see Fig.~\ref{fig:transition_cont}(b)]. Hence again we see that the $\cal PT$ transition here only depends on the absolute value of an eigenmode but not its phase angle, with which we can refer to it as a ``classical" transition similar to our discussion in Sec.~\ref{sec:PTdimer_CMT}.

\begin{figure}[b]
\centering
\includegraphics[clip,width=\linewidth]{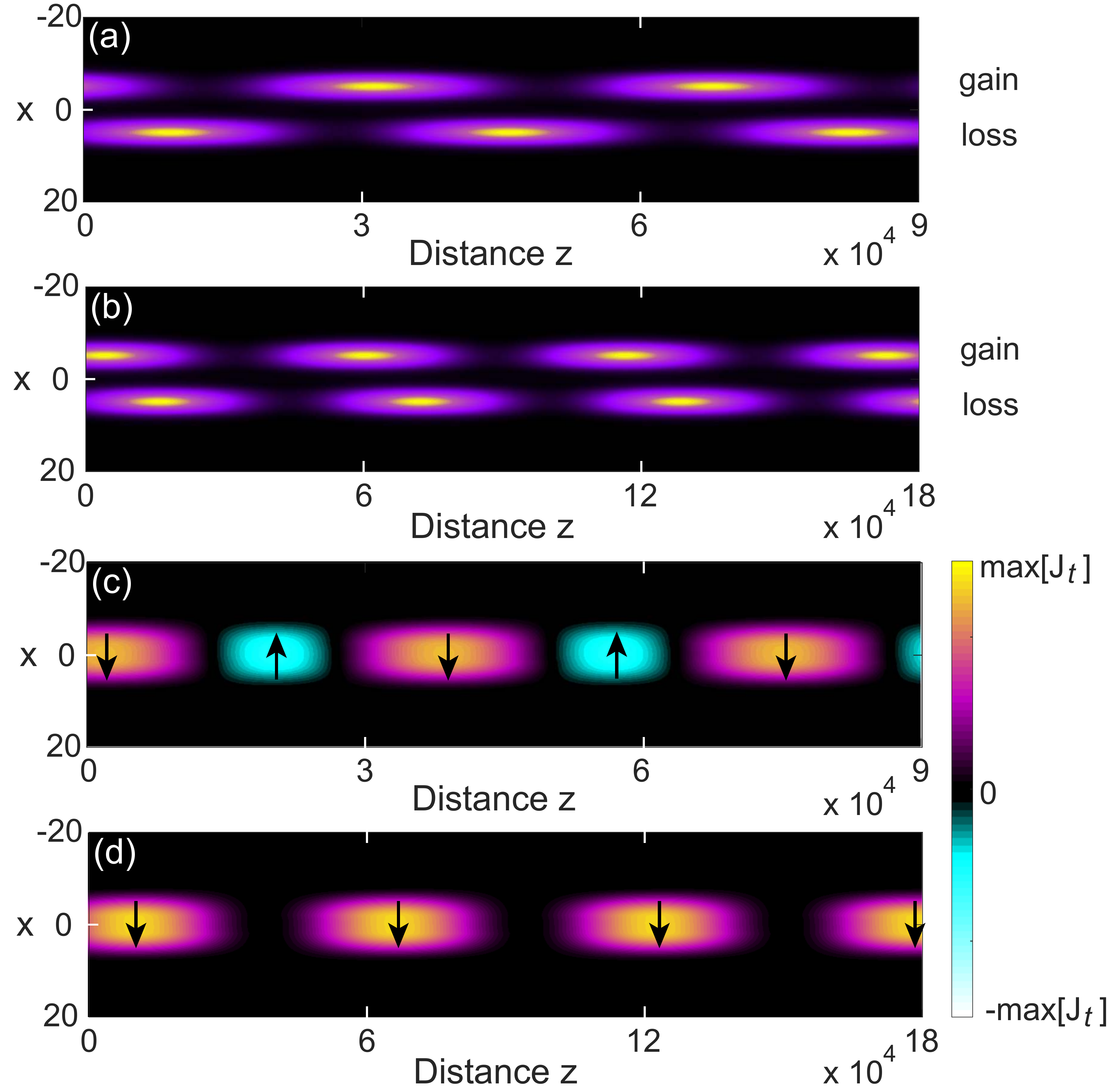}
\caption{(Color online) Power oscillation and transverse flux in the $\cal PT$ dimer shown in Fig.~\ref{fig:transition_cont}. (a)-(b) Wave intensities with $\tau=0.002$ and $0.005$, respectively. $c_+=2$, $c_-=1$ in Eq.~(\ref{eq:decomp2}). The colormap is the same as in Fig.~\ref{fig:osc}(a) and (b). (c) and (d) The corresponding transverse flux, which are bidirectional when $\tau=0.002$ and unidirectional when $\tau=0.005$. } \label{fig:osc1}
\end{figure}

Next we turn to power oscillations in the $\cal PT$-symmetric phase [see Figs.~\ref{fig:osc1}(a) and (b)] and discuss the transverse flux now defined by $J_t(x,z)=i[\Phi \partial_x\Phi^{*}-\Phi^{*} \partial_x\Phi]$. We first note that $J_t$ does not appear in Eq.~(\ref{eq:P_cont}), which is similar to the sum of Eqs.~(\ref{eq:I_a_pt}) and (\ref{eq:I_b_pt}) where $J_t$ is cancelled. In fact, the counterpart of Eqs.~(\ref{eq:I_a_pt}) and (\ref{eq:I_b_pt}) here is given by a continuity equation
\be
\frac{\partial |\Phi|^2}{\partial z} = - 2n_I(x)|\Phi|^2 + \frac{\partial J_t}{\partial x}, \label{eq:cont}
\ee
where the two terms on the right hand side represent intra-waveguide gain (or loss) and inter-waveguide flux exchange, just as in the coupled-mode theory.

In our $\cal PT$ dimer we define two waveguide regions of width $w$: $x\in[-d-w/2,-d+w/2]\equiv w_1$, $[d-w/2,d+w/2]\equiv w_2$, where $2d$ is the distance between their centers. For simplicity, we set $n_R(x\in w_{1,2})=-0.5$ and $n_I(x\in w_{1.2})=\pm\tau$, which are zero elsewhere.
To find an $x$-independent form of the transverse flux most relevant in the power oscillation phenomenon, we integrate Eq.~(\ref{eq:cont}) in the two waveguide regions, which result in
\begin{align}
\dz{I_1} &=  2\tau I_1 + J_t(-d+w/2,z),\\
\dz{I_2} &= -2\tau I_2 - J_t(d-w/2,z).
\end{align}
Here $I_{1,2}(z)\equiv\int_{w_{1,2}} dx|\Phi(x,z)|^2$ and we have taken $J_t(-d-w/2,z)=J_t(d+w/2,z)=0$ at the outer boundary of each waveguide [see Figs.~\ref{fig:osc1}(c) and (d)], where the waves of the guided modes are evanescent and do not carry a transverse flux. By comparing the two equations above to Eqs.~(\ref{eq:I_a_pt}) and (\ref{eq:I_b_pt}), their similarities are readily seen. Nevertheless, we note that $J_t(-d+w/2,z)\neq J_t(d-w/2,z)$, the difference of which equals the longitudinal flux in the inter-waveguide region $[-d+w/2,d-w/2]\equiv w_3$:
\be
\dz{(\int_{w_3}dx\,|\Phi|^2)} = J_t(d-w/2,z) - J_t(-d+w/2,z).
\ee
This difference, though small, can still be discerned from the slight skewness of $J_t$ shown in Figs.~\ref{fig:osc1}(c) and (d). More generally, we note
\be
J_t(-x,z)\neq J_t(x,z), \label{eq:flux_asymm}
\ee
which \textit{does not} contradict the following relation for an eigenmode in the $\cal PT$-symmetric phase:
\be
J_{t,n}(x)= J_{t,n}(-x). \label{eq:flux_symm_eig}
\ee
Equation (\ref{eq:flux_symm_eig}) is another manifestation of $\cal PT$ symmetry, which can be proven using $\Phi_n(-x) = \Phi_n^*(x)$ and $-\partial_x\Phi_n(-x) = \partial_x\Phi_n^*(x)$ in the $\cal PT$-symmetric phase. To understand the inequality (\ref{eq:flux_asymm}), we assume the two mode approximation
\be
\Phi(x,z)\approx c_+\phi_+(x)e^{-i\lambda_+z}+c_-\phi_-(x)e^{-i\lambda_-z}\label{eq:decomp2}
\ee
in our analysis, similar to Eq.~(\ref{eq:decomp}) in the coupled-mode theory. Again we choose $z=0$ to be where $c_{+,-}$ are in phase and denote $c_+=rc_-\,(r>0)$.
We then find $J_t(x,z) - J_t(-x,z) = 2r\left[(\phi_-\partial_x\phi_+^* - \phi_+^*\partial_x\phi_-)+c.c.\right]\sin(\beta_+-\beta_-)z$, which is zero only at the EP (where $\beta_+=\beta_-$).
Finally, we note that the transverse flux in the inter-waveguide region becomes unidirectional when $\tau$ is large [see Fig.~\ref{fig:osc1}(d)], just like our analysis based on the coupled-mode theory predicts.

\section{$\cal PT$ diamond}

\begin{figure}[t]
\centering
\includegraphics[clip,width=\linewidth]{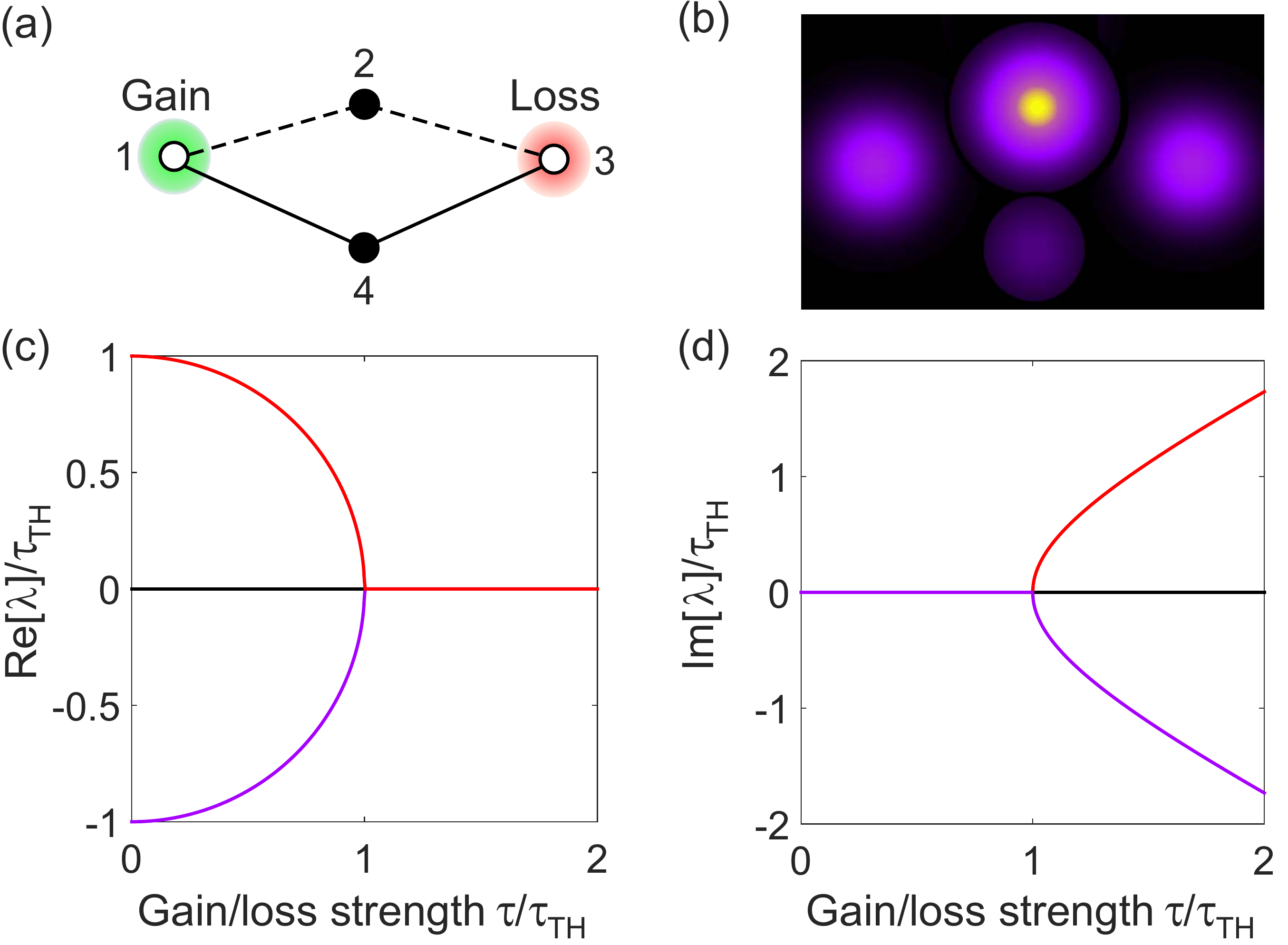}
\caption{(Color online) (a) Schematic of a $\cal PT$ diamond. Solid and dashed lines represent coupling strengths $J\in\mathbb{R}$ and $K=2J$, respectively. Filled and open dots show the two sublattices. The numbers label the positions of the waveguides in the wave functions. (b) Identical intensity pattern of the two modes with $\lambda\neq0$ in the $\cal PT$-symmetric phase. The color scheme is the same as in Figs.~\ref{fig:osc}(a) and \ref{fig:osc}(b). (c) and (d) Real and imaginary parts of the four eigenvalues as a function of the gain and loss strength. Two of them are always zero.} \label{fig:diamond}
\end{figure}

In this section we extend our examination of optical fluxes to a more complicated $\cal PT$-symmetric photonic system, which we refer to as a ``$\cal PT$ diamond." Instead of coupling the gain and loss waveguides directly, we now couple them indirectly via two separate waveguides without gain or loss [see Fig.~\ref{fig:diamond}(a)]. Having verified the validity of the coupled-mode theory, here we use it again in our analysis.

Below we focus on the shifted effective Hamiltonian $\tilde{H} = H - \beta\mathbb{1}$ of our $\cal PT$ diamond, where $\beta$ is the propagation constant of a single waveguide in Eqs.~(\ref{eq:a_pt}), (\ref{eq:b_pt}) and $\mathbb{1}$ is the identity matrix. Besides $\cal PT$ symmetry, $\tilde{H}$ also satisfies the non-Hermitian particle-hole (NHPH) symmetry, which is a universal property of gain and loss modulated photonic lattice with otherwise identical elements on two sublattices and with nearest neighbor coupling \cite{zeromodeLaser}. As a consequence of NHPH symmetry, the four eigenvalues of $\tilde{H}$ here satisfy
\be
\lambda_j = -\lambda_{j'}^*,
\ee
where $j,j'$ are not necessarily the same. When $j$ and $j'$ are different, $\lambda_{j,j'}$ are symmetric about the imaginary axis, and NHPH symmetry is spontaneously broken. 
When $j=j'$ instead, a ``zero mode" (i.e., $\re{\lambda_j}=0$) is formed on the imaginary axis of the complex $\lambda$ plane, where NHPH symmetry holds for the corresponding eigenstate.

For the four eigenmodes of the $\cal PT$ diamond, there is only one $\cal PT$-transition point [see Figs.~\ref{fig:diamond}(c) and (d)] which takes place at
\be
\tauEP=\sqrt{2(J^2+K^2)}.\label{eq:tauEP_diamond}
\ee
This $\cal PT$ transition is very similar to that of the $\cal PT$ dimer discussed previously: the trajectories of two $\lambda$'s form a semicircle in the $\tau-\re{\lambda}$ plane before the EP, with a radius given by $\tauEP$:
\be
\lambda_\pm = \pm\sqrt{\tauEP^2-\tau^2}\equiv\pm\tauEP\cos\phi,\quad\phi\in[0,\pi/2).
\ee
The same is true in the $\cal PT$ dimer, with $\tauEP=J$ instead. In addition, the wave functions of these two modes can be written as
\be
\Psi_\pm = \left[\frac{1}{\lambda_\pm-i\tau},\, \frac{K}{K^2+J^2},\, \frac{1}{\lambda_\pm+i\tau},\, \frac{J}{K^2+J^2}\right]^T. \label{eq:diamond_wf_pm}
\ee
Therefore, they have an identical intensity profile in the $\cal PT$-symmetric phase that \textit{do not} change with $\tau$, i.e.,
\be
|\Psi_\pm| = \left[\frac{1}{\tauEP},\, \frac{|K|}{K^2+J^2},\, \frac{1}{\tauEP},\, \frac{|J|}{K^2+J^2}\right]^T, \label{eq:diamond_int_pm}
\ee
which is shown in Fig.~\ref{fig:diamond}(b). The same behavior again takes place in the simple $\cal PT$ dimer, as can be seen directly from the expressions of $\Psi_\pm=[1,\pm e^{\mp i\phi}]$ given above Eqs.~(\ref{eq:I_a_pt2}) and (\ref{eq:I_b_pt2}).

The matrix rank of $\tilde{H}$ is 2 in this case, and hence the other two $\lambda$'s of $\tilde{H}$ ($H$) are always equal to 0 ($\beta$). Note however, $\lambda=0$ does not form an ``exceptional line" as a function of the gain and loss strength $\tau$, since these two modes are genuinely degenerate with distinct wave functions. One of them can be chosen as a ``dark state," i.e., one with zero amplitudes in the gain and loss waveguides:
\be
\Psi_1 = [0,\, J,\, 0,\, -K]^T \label{eq:diamond_wf_1}
\ee
It is then easy to understand why the eigenvalue of this mode is not affected by $\tau$ at all. If we require that the other wave function to be orthogonal with the dark state, it is then proportional to
\be
\Psi_2 = \left[\frac{1}{-i\tau},\, \frac{K}{K^2+J^2},\, \frac{1}{i\tau},\, \frac{J}{K^2+J^2}\right]^T \label{eq:diamond_wf_2}
\ee
and vary with $\tau$. It becomes an opposite dark state to $\Psi_1$ in the Hermitian limit, with vanishing amplitudes in the two bridging waveguides. Note that at the EP ($\lambda=0$) it becomes the same as the identical $\Psi_\pm$ given by Eq.~(\ref{eq:diamond_wf_pm}) [see Fig.~\ref{fig:diamond}(b)], and hence this EP is of order 3 \cite{Graefe}, similar to the finding in a 1D Lieb lattice with a flat band \cite{Flatband_PT}.

\begin{figure}[t]
\centering
\includegraphics[clip,width=\linewidth]{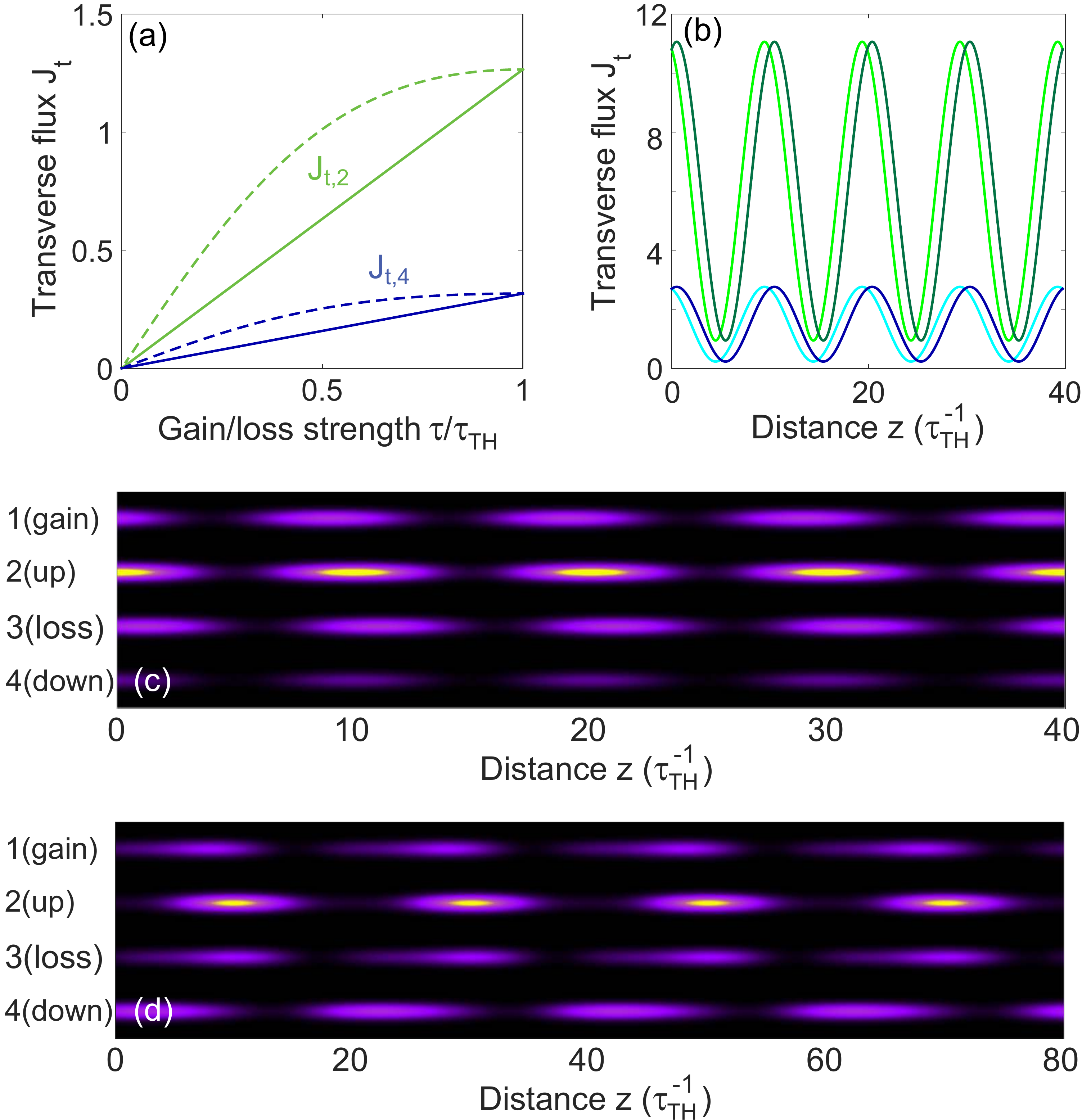}
\caption{(Color online) (a) Transverse fluxes in the eigenmodes of the $\cal PT$ diamond. The upper two and lower two curves show the fluxes through the upper and lower bridging waveguides [2 and 4 in Fig.~\ref{fig:diamond}(a)], with the solid lines for $\Psi_\pm$ (their fluxes are identical) and the dashed lines for $\Psi_2$. $K=2J$ as in Fig.~\ref{fig:diamond}. (b) Transverse fluxes when $\tau=3J$, $c_-=1$ and $r=2$ in Eq.~(\ref{eq:diamond_flux}), with the upper curves showing ${\cal J}_{t,12}$ (lighter green), ${\cal J}_{t,23}$ (darker green) and the lower curves showing ${\cal J}_{t,14}$ (lighter blue), ${\cal J}_{t,43}$ (darker blue). They are unidirectional in this case. (c) Power oscillations for the case shown in (b). (d) An example of generalized power oscillations when the input wave contains $\Psi_{1,2}$ also. The color scale in (c) and (d) is the same as in Figs.~\ref{fig:osc}(a) and (b).} \label{fig:diamond_flux}
\end{figure}

Now let us examine the transverse fluxes of our $\cal PT$ diamond in the $\cal PT$-symmetric phase. The first question we address is whether the fluxes through the two bridging waveguides are in the same direction (e.g., from the gain side to the loss side) for the three ``non-dark" eigenstates; the dark state obviously has zero fluxes due to the vanished wave amplitudes in the gain and loss waveguides. Since there is no gain or loss in the two bridging waveguides, they simple pass on the fluxes they receive from the gain (loss) waveguide to the other, i.e.,
\begin{align}
{\cal J}_{t,12} = iK(a^*\mu-a\mu^*) = {\cal J}_{t,23} = iK(\mu^*b-b^*\mu)&\equiv{\cal J}_{t,2} \nonumber\\
{\cal J}_{t,14} = iJ(a^*\nu-a\nu^*) = {\cal J}_{t,43} = iJ(\nu^*b-b^*\nu)&\equiv{\cal J}_{t,4} \nonumber.
\end{align}
$a,b$ are again the wave amplitudes in the gain and loss waveguides, and $\mu,\nu$ are those in the other two waveguides. The subscripts reflect the labeling shown in Fig.~\ref{fig:diamond}(a). Using the expressions for $\Psi_\pm$ and $\Psi_2$, we find that both ${\cal J}_{t,2}$ and ${\cal J}_{t,4}$ increase from zero as $\tau$ increases [see Fig.~\ref{fig:diamond_flux}(a)], which shows that indeed the gain waveguide serves as the ``source" in the $\cal PT$ diamond while the loss waveguide serves as the ``drain" along both pathways. This result is independent of the signs of the two coupling coefficients, because ${\cal J}_{t,2}\propto K^2$ and ${\cal J}_{t,4}\propto J^2$ hold for these modes. Below we will take $K$ and $J$ to be positive.

Next we focus on the power oscillations in the $\cal PT$-symmetric phase. For simplicity, we first consider an input wave that is a superposition of $\Psi_\pm$, similar to that in Eq.~(\ref{eq:decomp}). In this case ${\cal J}_{t,12},{\cal J}_{t,14}$ are again in the same direction, given by
\be
{\cal J}_{t,12}(z) = {\cal J}_0\{(1+r^2)\tau + 2r\tauEP\sin[2\phi_1(z)+\pi-\phi]\}\label{eq:diamond_flux}
\ee
and ${\cal J}_{t,14}(z) = J^2{\cal J}_{t,12}(z)/K^2$, where ${\cal J}_0={K^2|c_-|^2}/{\tauEP^2}$ and $\phi_1(z) = \tauEP z\cos\phi$. Here we have normalized the wave functions in Eq.~(\ref{eq:diamond_int_pm}) by the square root of its intensity. Different from the transverse fluxes in the eigenstates, now the two bridging waveguides no longer pass on the fluxes immediately; instead, they have a delayed response, i.e.,
\be
{\cal J}_{t,23}(z) = {\cal J}_0\{(1+r^2)\tau + 2r\tauEP\sin[2\phi_1(z)+\phi]\}\label{eq:diamond_flux2}
\ee
and ${\cal J}_{t,43}(z) = J^2{\cal J}_{t,23}(z)/K^2$.

Note that the four transverse fluxes given above have similar expressions as that in the simple $\cal PT$ dimer, and we again have unidirectional power transfer in the transverse direction [see Fig.~\ref{fig:diamond_flux}(b)] once the criterion given by Eq.~(\ref{eq:uni_criterion}) is satisfied [with $J=\tauEP$ there replaced by $\tauEP$ given by Eq.~(\ref{eq:tauEP_diamond})], despite the apparent oscillations of the intensity peak between the gain and loss waveguides [see Fig.~\ref{fig:diamond_flux}(c)].
We also note that since ${\cal J}_{t,14}(z), {\cal J}_{t,12}(z)$ are proportional and in phase [and so are ${\cal J}_{t,43}(z), {\cal J}_{t,23}(z)$], the power oscillations in the two bridging waveguides are always in phase [see Fig.~\ref{fig:diamond_flux}(c)], unlike those in the gain and loss waveguides.
The power oscillations in all four waveguides become more and more in phase as the system approaches the EP (not shown).

So far we have considered an input wave that is a superposition of $\Psi_\pm$. Below we briefly mention some properties of the wave dynamics if the input also contains $\Psi_{1,2}$. We note that power oscillations still exist between the gain and loss waveguides, in the sense that the same intensity pattern appears sequentially in them [see Fig.~\ref{fig:diamond_flux}(d)]; their dynamics are no longer described by a simple sinusoidal function of period $\pi/(\tauEP\cos\phi)$ as those reflected by Eqs.~(\ref{eq:diamond_flux}) and (\ref{eq:diamond_flux2}), but rather by a mixture of two harmonic components with periods $\pi/(\tauEP\cos\phi)$ and $2\pi/(\tauEP\cos\phi)$, with the latter giving the period of the combined dynamics. Also the dynamics in the two bridging waveguides are not in phase any more.

\section{Conclusion and discussions}

Using both a coupled-mode theory and the paraxial wave equation, we have revealed several surprising findings in $\cal PT$-symmetric photonic systems from the perspective of optical fluxes. In particular, we have shown that in a $\cal PT$ dimer the $\cal PT$ transition does not depend on the relative phase of the wave function in the two waveguides, and hence it can be considered as a classical effect. In addition, we have proved that the apparent power oscillation from the gain waveguide to the loss waveguide in the $\cal PT$-symmetric phase cannot be arbitrarily fast near the EP: while it is faster than that in the Hermitian case \cite{Bender_fast}, it is bounded by the inverse of the coupling constant. Furthermore, we have shown that it is always possible to have unidirectional transverse flux in the $\cal PT$-symmetric phase of a dimer, as long as the input is not a superposition of the two eigenmodes with equal amplitudes. We have found the same behavior in a $\cal PT$ diamond where the gain and loss waveguides are coupled via two bridging waveguides without gain or loss. We have also analyzed the ``giant amplification," a result of interfering two nearly identical eigenstates, and shown that it is a sub-exponential behavior and does not outperform a single waveguide with the same amount of gain.

We expect many of these findings to hold in certain more complicated $\cal PT$-symmetric photonic structures. For example, if we consider a one-dimensional lattice formed by coupling many identical $\cal PT$ dimers with the same coupling constant $J$, the forms of Eqs.~(\ref{eq:I_a_pt}) and (\ref{eq:I_b_pt}) remain the same with ${\cal J}_t$ replaced by ${\cal J}_t+{\cal J}_{t2}$, where ${\cal J}_{t2}\equiv iJ(a^*be^{-ik\Lambda}-ab^*e^{ik\Lambda})$, $k$ is the wave vector, and $\Lambda$ is the lattice constant. Since ${\cal J}_t$ does not affect $\im{\lambda}$ [and again $\im{\lambda}={\cal J}_z/2(I_a+I_b)$], the $\cal PT$ transition again only depends on whether $I_a=I_b$ and hence can still be treated as a classical effect. Similarly, the peak-to-peak distances $z_{1,2}$ are given by the same expressions as Eqs.~(\ref{eq:z1}),(\ref{eq:z2}) with the coupling $J$ replaced by an effective coupling $\tilde{J}=\sqrt{2J(1+\cos k\Lambda)}$, the inverse of which now limits how fast the apparent power transfer from a gain waveguide to its two neighboring loss waveguides can take place. Although the expressions for ${\cal J}_t$ (and ${\cal J}_{t2}$) becomes more complicated, one can show that for a large enough $\tau$ in the $\cal PT$-symmetric phase of a given $k$, the transverse flux is always ``unidirectional," i.e., only from the gain waveguides to the loss waveguides, as long as $r\neq1$ for the wave functions $\Psi_\pm(k)$ in the two bands.

L.G. acknowledges support by NSF under Grant No. DMR-1506987. K.G.M. acknowledges support by the European Commission under projects NOLACOME (PIOF
303228), NHQWAVE (MSCA-RISE 691209), and by the European Union Seventh Framework Program (FP7-REGPOT-2012-2013-1) under grant agreement 316165.

\bibliographystyle{longbibliography}

\end{document}